\documentstyle[emulateapj]{article}

\lefthead{Armitage \& Livio}
\righthead{Black hole formation during common envelope evolution}

\begin{document}

\submitted{ApJ submitted 5/17/99}

\title{Black hole formation via hypercritical accretion during common 
       envelope evolution}

\author{Philip J. Armitage\altaffilmark{1} and Mario Livio}
\affil{Space Telescope Science Institute, 3700 San Martin Drive,
        Baltimore, MD 21218; \\
        armitage@cita.utoronto.ca, mlivio@stsci.edu}
\altaffiltext{1}{Current address: 
       Canadian Institute for Theoretical Astrophysics, McLennan Labs, 
       60 St George St, Toronto, M5S 3H8, Canada}     

\begin{abstract}
Neutron stars inspiralling into a stellar envelope can accrete at rates vastly 
exceeding the Eddington limit if the flow develops pressures high enough 
to allow neutrinos to radiate the released gravitational energy. It has 
been suggested that this hypercritical mode of accretion leads 
inevitably to the formation of stellar mass black holes during common 
envelope evolution. We study the hydrodynamics of this flow at 
large radii ($R \gg R_{\rm ns}$), and show that for low Mach number 
flows, in two dimensions, modest density gradients in the stellar envelope 
suffice to produce a hot, advection dominated accretion disk around the 
accreting object. The formation of outflows from such a disk is 
highly probable, and we discuss the impact of the resultant mass loss
and feedback of energy into the envelope for the survival of the 
neutron star. Unless outflows are weaker than those inferred for 
well observed accreting systems, we argue that in most cases insufficient accretion 
occurs to force collapse to a black hole before the envelope has been ejected.
This conclusions is of interest for black hole formation in general, for 
some models of gamma ray bursts, and for predictions of the event rate 
in future LIGO observations.
\end{abstract}	


\section{INTRODUCTION}

Common envelope evolution, in which the components of a binary system 
are engulfed by a common gaseous envelope, is a brief but crucial 
phase in the formation of many compact binary systems. Drag forces, due 
to velocity differences between the orbiting components of the 
binary and the surrounding gas, work to shrink the binary orbit, while the 
potential energy released acts to expel the common envelope. Extensive 
numerical work (e.g. Bodenheimer \& Taam 1984; Livio \& Soker 1988; Taam,
Bodenheimer \& Rozyczka 1994; Sandquist et al. 1998) has only partially succeeded in 
reducing the large uncertainties surrounding the efficiency of this 
process (Iben \& Livio 1993; Livio 1994; Rasio \& Livio 1996). 

For neutron stars, an equally fundamental uncertainty concerns 
the {\it accretion rate} during the common envelope phase. At the 
densities of a typical giant envelope, the 
Bondi-Hoyle accretion rate $\dot{M}_{\rm BH}$ for an inspiralling neutron star 
would be extremely large, typically of the order of 
$\sim 1 \ M_\odot {\rm yr}^{-1}$, and often much larger still. 
This is many orders of magnitude 
in excess of the Eddington rate\footnote{the Eddington `limit' here 
implies only an order-of-magnitude estimate of the accretion rate, since 
accretion in common envelopes is not generally spherically symmetric} 
of $\sim 10^{-8} \ M_\odot {\rm yr}^{-1}$ --- obtained by equating the 
outward radiation pressure with gravity --- leading to starkly different 
predicted outcomes. If $\dot{M}$ is limited to the Eddington rate (or 
is even several orders of magnitude higher), 
then ejection of the stellar envelope will occur when the total accreted 
mass is $M_{\rm acc} \ll M_\odot$, and the neutron star will survive. 
Conversely, uninhibited accretion at close to $\dot{M}_{\rm BH}$ will lead 
almost inevitably to collapse to a black hole. 

Recent work has tended to suggest that $\dot{M}_{\rm BH}$ is the 
appropriate rate for neutron star accretion in a stellar envelope
(see e.g. Bethe \& Brown 1998 and references therein). At 
sufficiently high accretion rates, the Eddington limit becomes irrelevant 
because photons are trapped and advected inward with the flow (Rees 1978; 
Begelman 1979; Blondin 1986). A hot 
envelope develops around the neutron star, and eventually high enough densities 
and temperatures are reached that the accretion energy can be radiated 
away by neutrinos. This {\it hypercritical} mode of accretion has been considered 
in the context of fallback onto neutron stars in supernovae (e.g. Colgate 1971; 
Zeldovich, Ivanova \& Nadezhin 1971; Chevalier 1989; Houck \& Chevalier 1991), 
as a component of models for gamma-ray bursts (Popham, Woosley \& Fryer 1999), 
and applied to the problem of accretion in a stellar envelope (Chevalier 1993, 
1996; Brown 1995; Fryer, Benz \& Herant 1996). A striking consequence of 
ubiquitous hypercritical accretion would be that neutron star-neutron star 
binaries {\it cannot} form from massive binaries via the usual common 
envelope evolution route, which would instead lead to black hole-neutron 
star binaries. The observed population of neutron star-neutron star binaries 
would have to form instead via a rarer channel in which the binary progenitors 
are of nearly equal mass (Brown 1995), and would be an order of magnitude rarer 
than black hole-neutron star binaries (Bethe \& Brown 1998). The robustness of 
this conclusion is of great interest, since these binaries are both potential 
gamma-ray burst progenitors, and the most promising targets for the early 
LIGO detection of gravitational waves (Abramovici et al. 1992).

The velocity of the neutron star relative to the common envelope ensures 
that the accretion flow cannot be spherically symmetric, at least at 
large radii of the order of the accretion radius. Moreover, gradients 
in the envelope structure across the accretion radius introduce 
angular momentum into the flow, which modifies its properties even 
at small radii, via the formation of a rotationally supported disk. 
In this paper we investigate how these complexities affect hypercritical 
accretion via hydrodynamic simulations of the outer parts of the 
accretion flow. Our approach thus complements the consideration of angular 
momentum presented by Chevalier (1996), though as discussed later 
we draw different conclusions as to its likely importance.

The plan of this paper is as follows. In \S2 we briefly review the 
properties of spherical hypercritical accretion that are relevant for the 
common envelope application, and in \S3 we outline the numerical 
methods used. The principal computational limitation is the 
restriction to two dimensional simulations. Results for flows 
with zero and non-zero angular momentum are presented in \S4 and 
\S5, and the implications and remaining uncertainties discussed 
in \S6.

\section{HYPERCRITICAL ACCRETION}

The simplest approximation to the accretion flow in a common 
envelope is to describe it as Bondi-Hoyle-Lyttleton accretion 
from a uniform medium onto a point mass. The properties of 
such flows have been extensively studied, both analytically 
(Hoyle \& Lyttleton 1939; Bondi \& Hoyle 1944), and numerically 
(e.g. Ruffert 1997; Kley, Shankar \& Burkert 1995; 
Benensohn, Lamb \& Taam 1997, and references therein). For an 
accretor of mass $M$, the critical radius is the accretion 
radius\footnote{There are a variety of definitions of $R_a$ in 
the literature, but all lead to accretion rates that are 
essentially the same for our purposes}, 
\begin{equation} 
  R_a = { {2GM} \over {v_\infty^2 + c_\infty^2} },
\label{eq1}
\end{equation}   
where $v_\infty$ and $c_\infty$ are the velocity and sound speed 
far upstream of the accreting star. Material falling into 
a cylinder with this radius is accreted, so that, 
\begin{equation}
  \dot{M}_{\rm BH} = \pi R_a^2 \rho_\infty \sqrt{v_\infty^2 + c_\infty^2},
\label{eq2}
\end{equation}  
where $\rho_\infty$ is the density of the gas far upstream of the 
accretor, and the last factor is included as it provides better results 
for low Mach number accretors. The timescale for setting up the flow 
is of the order of the sound crossing time at the accretion 
radius, $t_a = R_a / c_\infty$. Generally, since we are interested 
in inspiral through a pressure supported atmosphere, we expect 
Mach numbers of order unity, though this will vary somewhat at different 
radii in the star. 

Numerical values for $R_a$ and $\dot{M}_{\rm BH}$ require specification 
of a stellar model. Fryer, Benz \& Herant (1996) tabulate values for 
several giant and main sequence stars (the giants are of greater interest here, 
though somewhat analagous considerations apply to neutron star-main sequence 
collisions, for example in globular clusters). For example, for a 
$20 \ M_\odot$ giant $R_a \sim 10^{11} \ \rm{cm}$, while $\dot{M}_{\rm BH}$ 
varies from $10^2 \ M_\odot {\rm yr}^{-1}$ on upwards. For a more 
distended $10 \ M_\odot$ model lower accretion rates of $10^{-2} \ M_\odot 
{\rm yr}^{-1}$ are of course expected in the outer regions.
These accretion rates can only be realized in the hypercritical 
regime, since the Eddington limit on the 
luminosity,
\begin{equation} 
  L_{\rm Edd} = { { 4 \pi G M m_p c } \over {\sigma_T} },
\label{eq3}
\end{equation} 
where $m_p$ is the proton mass and $\sigma_T$ the Thomson cross-section, 
corresponds to a few $\times 10^{-8} \ M_\odot {\rm yr}^{-1}$.

For hypercritical accretion to occur the timescale 
for photons to diffuse out of the flow must exceed the timescale for them 
to be advected inwards. For a spherical flow whose opacity $\kappa_{\rm es}$ 
is dominated by electron scattering, the optical depth is 
\begin{equation} 
  \tau_{\rm es} = \int \rho(R) \kappa_{\rm es} dR, 
\label{eq4}
\end{equation}
and the diffusion time is roughly,
\begin{equation} 
  t_{\rm diff} \approx { R \over c} \tau_{\rm es}.
\label{eq5}
\end{equation}
For a flow in free-fall, where $v_r = - \sqrt{2 G M / R}$, equating $t_{\rm diff}$ to 
the free-fall time $R / v_r$ yields an estimate for the trapping radius
(Blondin 1986),
\begin{equation}
  R_{\rm trap} = {1 \over 2} { {\dot{M} c^2} \over L_{\rm Edd} } { {2 G M} \over c^2 }.
\label{eq6}
\end{equation}  
In addition to the photons being trapped, we must also be able to eventually 
lose the accretion energy from the system. For a black hole, this is possible 
at any accretion rate since the energy can be invisibly advected across the 
horizon (e.g. Narayan, Garcia \& McClintock 1997). In a neutron star system, 
high accretion rates are required because the pressure at the base of the 
accretion flow must be able to become large enough to allow neutrino losses 
(primarily from electron positron annihilation into neutrino pairs) to balance 
the accretion energy, $GM \dot{M} / R_{\rm ns}$. In the spherically 
symmetric case, this requirement usually requires the formation of 
a hot envelope at small radii, in which $v_r$ is subsonic, bounded 
from free falling gas by a shock at radius $R_{\rm sh}$. Within this inner 
region, the density and pressure scale as a radiation dominated 
$\gamma = 4/3$ ($n=3$) polytrope (e.g. Brown 1995; Fryer, 
Benz \& Herant 1996), 
\begin{equation} 
  \rho = \rho_0 \left( R \over R_{\rm sh} \right)^{-3}, \ \ \ \ 
  p = p_0 \left( R \over R_{\rm sh} \right)^{-4}. 
\label{eq7}
\end{equation}  
A detailed calculation (Houck \& Chevalier 1991; Brown 1995) finds 
for $R_{\rm sh}$,
\begin{equation} 
  R_{\rm sh} \simeq 2.6 \times 10^{8} \ {\rm cm} 
  \left( \dot{M} \over { M_\odot \ {\rm yr}^{-1} } \right)^{-0.37}.
\label{eq8}
\end{equation}  
Requiring that $R_{\rm sh} < R_{\rm trap}$ gives a limit on the 
accretion rate, of roughly 
$\dot{M}_{\rm hyper} \gtrsim 10^{-4} \ M_\odot {\rm yr}^{-1}$. Even 
allowing for a generous margin of error in this crude estimate, for example 
due to a non-spherical geometry, this is substantially less than the 
Bondi-Hoyle rate in an envelope, which should thus be safely in the 
hypercritical regime. Of course we also require in the common 
envelope case that $R_{\rm sh} < R_a$, but for spherical accretion 
and $R_a \sim 10^{11} \ {\rm cm}$ this is an easily satisfied 
constraint, $\dot{M} \gtrsim 10^{-7} \ M_\odot {\rm yr}^{-1}$.

If angular momentum and other possible complications can be neglected, 
the scenario for the inspiral of a neutron star then has three stages. 
First, the neutron star accretes at less than the Eddington rate as 
it encounters the tenuous outer atmosphere of the companion. Negligible 
mass is accumulated in this phase. As the density grows, the implied 
Bondi-Hoyle rate first rises above $\dot{M}_{\rm Edd}$, but is below 
$\dot{M}_{\rm hyper}$. In this regime outward diffusion of photons is 
effective at limiting accretion, again to negligible levels. Finally, 
the Bondi-Hoyle rate rises well above the critical value for hypercritical 
accretion, and gas starts accreting freely onto the neutron star. The 
outcome then depends critically on how much mass can be accumulated 
before enough energy has been lost to unbind the envelope. Brown (1995)
estimates this energy as, 
\begin{equation} 
 E \simeq 3 < v_\infty^2 > \Delta M
\label{eq9}
\end{equation}
where $\Delta M$ is the accreted mass, and $<v_\infty^2>$ is the average  
mean velocity squared during the hypercritical accretion phase. For 
typical binding energies and velocities, $\Delta M \gtrsim 1 \ M_\odot$, 
sometimes vastly so, and the mass accreted will probably exceed the 
maximum mass for a neutron star and force collapse to a black hole.

\section{NUMERICAL METHOD}

We investigate how angular momentum could affect the above scenario 
of hypercritical accretion by studying the inviscid, purely 
hydrodynamic behavior of infalling gas at large radii, $R \gg R_{ns}$. 
The hypercritical regime corresponds to the complete dominance 
of advection of photons over diffusion, so that the effects of 
radiation transport are negligible. We also neglect magnetic 
fields, which is potentially less forgivable. To attain the 
required dynamic range in three dimensional 
calculations including magnetic fields unfortunately remain difficult.

\subsection{2D simulations}

We have investigated two dimensional simulations both in spherical polar 
geometry (with $v$ parallel to $\theta = 0$ and axisymmetry 
in the $\phi$ direction), and in cylindrical geometry 
$(z,R,\phi)$. The spherical polar simulations represent 
the most faithful two dimensional representation of the 
flow, but necessarily exclude the possibility of studying 
angular momentum or density gradients in the ambient medium. 
For such zero angular momentum flows, we found only that 
the analytic Bondi-Hoyle estimate of equation (\ref{eq2}) 
provides a good estimate of the accretion rate (for low 
Mach numbers the actual accretion rate is somewhat 
higher than the estimate, but the difference is not 
so great as to qualitatively affect the result).

For these reasons (and following Benensohn, Lamb \& Taam 1997 
and most other two dimensional calculations) we focus on the 
cylindrical polar calculations. The neutron star is represented 
as a point mass at $R=0$, surrounded by a fixed computational mesh in 
$(R,\phi)$. We use uniform zoning in the $\phi$ direction, and choose
the radial grid such that $R_{i+1} = \beta R_i$ with $\beta > 1$ a 
constant. This amounts to choosing cells that have the same 
shape, in our case roughly square, at all radii. 

Cylindrical polar simulations in which $z$ is the ignored 
co-ordinate allow for angular momentum, but correspond 
to the rather unphysical assumption that the central 
mass moves through a thin sheet of gas with zero gradients 
in the perpendicular direction. This may lead to qualitative 
errors, for example in the strength and prominence of 
`flip-flop' instabilities at high Mach numbers (compare e.g. 
Livio 1992; Benensohn, Lamb \& Taam (1997); Ruffert (1997)), 
though untangling the effects of geometry and resolution in 
the various calculations is difficult. We will not be 
interested here in transient disks or instabilities in 
the wake, where these issues are most worrisome.

We assume that the gas is dominated by radiation pressure out 
to the outer boundary of the simulation at a few $R_a$. The 
equation of state can then be modelled using a simple 
adiabatic relation, 
\begin{equation} 
  p = (\gamma - 1) \epsilon, 
\label{eq10}
\end{equation} 
where $\epsilon$ is the energy per unit volume and we take 
$\gamma = 4/3$, corresponding to a radiation dominated gas 
where $p=1/3 \epsilon$.  

The calculations use the ZEUS-3D code developed by the 
Laboratory for Computational Astrophysics (Clarke, Norman \& 
Fiedler 1994). ZEUS is an Eulerian finite difference code 
which employs an artificial viscosity to handle shocks. 
The algorithms and design of the code are closely similar 
to those detailed by Stone \& Norman (1992a, 1992b).

\subsection{Boundary conditions}

For the outer boundary condition, we impose inflow of gas 
at velocity $v_\infty$ and sound speed $c_\infty$ at the 
outer boundary, $R_{\rm out}$, for $-\pi / 2 < \phi < \pi / 2$. 
Over the remainder of the outer boundary outflow boundary 
conditions are specified, implemented as simple continuation of fluid 
variables on the grid into the boundary zones. We take a Mach 
number ${\cal{M}} = 1.5$, appropriate to the generally mildly 
supersonic accretion flows in common envelopes.

An inspiralling neutron star encounters radial gradients in 
$\rho_\infty$, $c_\infty$ and $v_\infty$ across the accretion 
radius. In most cases the density gradient is the principal 
effect (e.g. Fryer, Benz \& Herant 1996), which we model 
as a simple exponential,
\begin{equation} 
  \rho_\infty \propto e^{ \epsilon_\rho \Delta r / R_a }
\label{eq11}
\end{equation}
where $\Delta r$ is the radial distance in the common envelope 
of the unperturbed medium from the neutron star orbit, and 
$\epsilon_\rho$ measures the strength of the gradient across 
the accretion radius. Values of $\epsilon_\rho$ for various 
stars vary substantially, and have been tabulated by Fryer, 
Benz \& Herant (1996).

The inner boundary condition is reflecting ($v_R = 0$ at $R=R_{\rm in}$), 
allowing for the formation of a pressure supported inner envelope 
for zero angular momentum accretion. This implicitly assumes 
low accretion rates where the shock radius $R_{\rm sh}$ is large 
enough to exceed $R_{\rm in}$. For runs with angular momentum, 
the choice of inner boundary condition is less important as 
angular momentum provides substantial support against gravity 
at small radii. Our boundary condition then amounts to assuming 
that radial flow through the disk is slow -- the plausibility of 
this can be verified post facto by studying the properties of 
the disks formed in the calculation.

\section{DISK FORMATION}

Figure 1 shows results for a series of Bondi-Hoyle accretion 
simulations in which the density gradient in the ambient 
medium was varied from $\epsilon_\rho = 0$ to $\epsilon_\rho = 0.4$. 
All the calculations used a grid with $n_\phi = 144$ and 
$n_R = 160$. The inner boundary was at $R_{\rm in} = R_a / 60$ and the outer 
boundary at $R_{\rm out} = 4 R_a$, giving a grid with $\beta \simeq 1.03$.
The calculations were run until $t = 32 t_a$, where the time unit 
$t_a$ is the sound crossing time of the accretion radius $R_a$. We 
plot in Fig.~1 only the inner region of the accretion flow. The 
$\epsilon_\rho = 0.2$ run was also recomputed at modestly higher 
resolution ($n_r = n_\phi = 200$) until $t = 100 t_a$, in order 
to ascertain how quickly mass accumulated in the outer parts of the 
disk at a later time. No qualitative changes were observed to occur during 
this longer simulation.

In the absence of density gradients in the ambient gas, the structure 
of the flow resembles closely that seen in simulations of Bondi-Hoyle 
accretion with the same parameters and an absorbing inner boundary condition.
A pressure supported, roughly symmetric envelope has developed around 
the central object, and this displaces the bow shock upstream into 
the flow. At this resolution and Mach number, the flow is found to 
be only rather weakly transient. Increasing the density gradient 
first leads to an asymmetric displacement of the bow shock, followed 
for more extreme density gradients by the formation of a clear 
disk in the inner regions of the simulation. For the parameters 
adopted here, $\epsilon_\rho \gtrsim 0.2$ suffices to create a
clear and persistent disk surrounding the central object.

Figure 2 plots the velocity and density fields corresponding to 
the images in Fig.~1. Over this radial range there is a large variation 
in velocity, which we normalise to the local Keplerian value 
$v_k = \sqrt{(GM / R)}$. For the zero density gradient case, the outer flow 
is clearly not cylindrically symmetric, but within $\sim 0.1 R_a$ of the 
central object the envelope is both reasonably symmetric and characterised 
by only small rotational motions. Conversely, a clear disk is 
produced in the calculations with stronger density gradients, 
$\epsilon_\rho \gtrsim 0.2$, the flow here is disk-like out to 
at least $R_a / 2$.

Figure 3 shows how the volume averaged rotational velocity from the 
simulations, normalised to $v_k$, varies as a function of radius. 
Negligible rotation is seen in the $\epsilon_\rho = 0$ run, with 
$<v_\phi> / v_k \leq 0.1$ at all radii, while the runs with 
density gradients all show the signature of a disk in which 
both pressure gradients and rotational support are significant. 
For the two runs with the largest gradients, a sizeable disk 
extending out to almost $R_a$, with typical angular velocity 
$\Omega \approx \Omega_k / 2$, is clearly produced.

\section{DISK EVOLUTION}

For hypercritical accretion, the disks formed around the 
neutron star will be advection dominated. The properties 
of advection dominated flows have been extensively studied, 
both at the high accretion rates of interest here (e.g. Begelman \& 
Meier 1982) and more recently at lower accretion rates (e.g. Narayan \& 
Yi 1994, 1995). These disks are hot, geometrically thick, and only
weakly bound to the accreting object. The presence of a disk 
may affect the outcome of common envelope evolution in two 
ways, via a modification of the accretion rate onto the neuton star 
or via feedback of energy into the stellar envelope. 

The most direct potential influence of a disk is via the 
accretion rate. As infall continues, the disk 
will reach a quasi-steady state in which the rate of infall onto 
the outer disk balances the rate of disk accretion. This 
rate need not be the Bondi-Hoyle rate, and in principle could be  
lower (for an illustration, in our inviscid simulations, 
where the viscosity is very low, we eventually reach a 
steady state where there 
is close to {\em zero} ongoing accumulation of mass in the disk). 
We estimate whether this is important below, and show that for 
the usually considered values of $\alpha_{\rm SS}$, the 
Shakura-Sunyaev (1973) viscosity parameter, the disk is 
probably able to transport mass inwards at the Bondi-Hoyle 
rate.

The presence of a disk also makes the formation of outflows 
or jets probable (e.g. Livio 1999). Strong outflows could themselves affect 
the fate of the system by reducing the accretion rate 
onto the neutron star. Much weaker jets, if they arose from deep in 
the neutron star potential well, would still 
provide an important additional energy input into the 
envelope, and shorten the common envelope phase.

\subsection{Accretion rate}

The accretion rate through the disk can be estimated from the 
measured disk mass and the inferred viscous timescale. For a 
viscosity parameterized via the Shakura-Sunyeav (1973) form, 
$\nu = \alpha_{\rm SS} c_s^2 / \Omega_k$, and at radius $R$ 
the viscous timescale is,
\begin{equation} 
  t_\nu = { R^2 \over \nu } = { {R^2 \Omega_k} \over {\alpha_{\rm SS} 
  c_s^2} }.
\label{eq11b}
\end{equation}
The mass in the disk will drain onto the central object on the viscous 
timescale at the outer radius, with an accretion rate 
$\dot{M}_{\rm disk} \approx M_{\rm disk} / t_\nu$. The value 
of $\alpha_{\rm SS}$ appropriate to thick, radiation dominated 
disks is unknown. If MHD instabilites (Balbus \& Hawley 1991) are 
responsible for the viscosity, then simulations of thin gas pressure 
dominated disks find typically that $\alpha_{\rm SS} \approx 10^{-2}$ 
(Stone et al. 1996; Brandenburg et al. 1996). 
There remain large uncertainties in the 
theoretical expectation for disks of the kind that we are interested in.
However, taking $\alpha_{\rm SS} = 10^{-2}$, and evaluating the accretion 
rate from the disk mass and run of sound speed obtained at $t=100 t_a$ in the 
long duration $\epsilon_\rho = 0.2$ run, we find that $\dot{M}_{\rm disk} \sim 
\dot{M}_{\rm BH}$. This is a crude estimate, which in particular ignores 
completely the expansion of the disk expected from viscous evolution 
(Lynden-Bell \& Pringle 1974). As a result, the value $\dot{M}_{\rm BH}$ 
should be regarded as an upper limit to the mean accretion rate through 
the disk.
Furthermore, at least at small radii the 
accretion is unlikely even to be steady (Chevalier 1996), since 
the much slower fall-off of pressure with radius in an 
advection dominated disk as compared to a spherical 
envelope ($p \propto R^{-5/2}$ as compared to $p \propto R^{-4}$) 
requires non-steady accretion to reach the extreme conditions 
required for neutrino emission.
However, it suggests that there is 
no strong reason to believe that, in the absence of outflows, the formation of a disk
creates an insurmountable bottleneck to rapid accretion. 

\subsection{Outflows and jets}

Most accretion disk systems are observed to 
generate jets or less well collimated outflows. Although 
a thick disk such as that formed in the simulations has 
a relatively small density contrast between the equatorial 
plane and the polar regions, advection dominated flows 
are also hot enough that the gas is only weakly bound 
to the central mass. Detailed solutions show that the 
Bernoulli constant, 
\begin{equation}
  {\rm Be} \equiv {1 \over 2} v_R^2 + {1 \over 2} \Omega^2 R^2 
  - \Omega_k^2 R^2 + { \gamma \over {\gamma - 1} } c_s^2,
\label{eq12} 
\end{equation}
which measures the energy the gas would possess if adiabatically 
moved to infinity, is positive for an often wide range of 
angles close to the poles (Narayan \& Yi 1994; 1995). Although 
the outcome depends additionally on the outer boundary 
conditions and the detailed physics of outflow generation, 
this positivity of ${\rm Be}$ is likely to imply 
that outflows are a generic feature of advection dominated 
disks ( Narayan \& Yi 1995; Blandford \& Begelman 1999). For 
our purposes, we can distinguish two extreme possibilities, 
in which outflows are either self-similar or generated 
exclusively from the inner disk at $R \simeq R_{\rm ns}$.

\subsubsection{Outflows}

{\em Self-similar} outflows from advection dominated 
flows represent the model considered by Blandford \& Begelman (1999). 
In this case the fraction of accreting mass lost in the outflow 
is the same for each decade in disk radius, so that the remaining mass
accretion rate through the disk decreases inwards as $\dot{M} \propto R^n$ 
with $0 < n < 1$. The appropriate value of $n$, which within this model measures the 
efficiency with which mass is ejected, has not been determined for any 
realistic thick disk model, though it is has been suggested that $n$ is 
large for simulations of convection in thick 
disks (Stone, Pringle \& Begelman 1999). Empirically $n$ must 
be $n \sim 1$ if this model is to be successful in explaining 
the extremely low radiative efficiencies of disks around black holes in 
low luminosity galactic nuclei (e.g. Reynolds et al. 1996). 
In the common envelope case, the disk extends over an extremely 
large range of radii, from the neutron star surface at $R \sim 10^6 \ {\rm cm}$ 
out to of the order of the accretion radius at $R \sim 10^{11} \ {\rm cm}$. As 
a result, the consequence of outflows that are inefficient at removing 
angular momentum would be to greatly reduce 
the accretion rate onto the neutron star. Cygnus X-2 may be an 
example of a neutron star that has survived an episode of super-Eddington 
mass transfer (though here during thermal timescale mass transfer rather 
than common envelope evolution) without accreting a significant fraction 
of the transferred mass (King \& Ritter 1999; King \& Begelman 1999). 
If this is indeed the case, it provides some support for the scenario of 
strong outflows under physical conditions analagous to those encountered during 
common envelope inspiral.

\subsubsection{Jets}

Alternatively, outflows may arise predominantly from the 
{\em inner disk}. In this case, the energy feedback into the 
common envelope from outflows originating deep in the neutron 
star potential well could be highly significant. If, as 
observations of jet systems suggest (Livio 1999), the 
jet is launched with a velocity roughly equal to the local 
Keplerian velocity, $v_{\rm jet}^2 \sim GM_{\rm ns} / R_{\rm ns}$, 
then the energy deposition after $\Delta M_{\rm jet}$ of mass has 
been ejected is just,
\begin{equation}
  E_{\rm jet} \simeq \alpha_{\rm jet} { {G M_{\rm ns} \Delta M_{\rm jet} } 
  \over R_{\rm ns} },
\label{eq13}
\end{equation}   
where $\alpha_{\rm jet}$ is an efficiency factor that will depend on 
the specific jet model. Ejecting the common envelope requires an 
energy deposition of around $2 \times 10^{48} \ {\rm ergs}$ (Brown 1995), 
which could be achieved for a mass loss in a jet as low as,
\begin{equation} 
  \Delta M_{\rm jet} \simeq 5 \times 10^{-6} \alpha_{\rm jet}^{-1} 
  \ M_\odot.
\label{eq14}
\end{equation}
The energetic feedback from such an outflow could thus be important 
for hastening the ejection of the envelope, even if the mass loss 
itself was far too small to significantly impact the accretion rate.
Indeed for this type of jet a large energy deposition 
could only be avoided if either the accretion rate 
was $\ll \dot{M}_{\rm BH}$, or if the efficiency of the jet production 
or coupling to the envelope was extremely low. A low efficiency of 
coupling to the envelope could arise if the jet was extremely 
well collimated and able to escape the star entirely, which remains 
a possibility. However, even for a collimation angle of 
$\sim 10^{-2}$, which is typical for many jet sources, the ejection 
of $\sim 0.2 M_\odot$ is sufficient to unbind the envelope. 
If jets are able to deposit energy into the envelope then 
an immediate consequence would be 
that the outcome of common envelope evolution should depend on 
the depth of the potential well at the surface of the inspiralling 
compact object -- more compact objects should lead to a higher value of  
the common envelope efficiency parameter 
$\alpha_{\rm CE}$ and eject the envelope more easily.    

\section{DISCUSSION}

In this paper we have considered the probable fate of neutron 
stars during common envelope evolution. Numerical simulations 
of the hydrodynamics of Bondi-Hoyle accretion at large 
radii from the neutron star show that modest density gradients, 
typical of those expected in giant envelopes, lead to persistent 
rotationally supported disks around the accreting object. The 
presence of angular momentum means that spherically symmetric 
treatments of hypercritical accretion are likely to be a poor 
approximation for studying the the common envelope phase. In 
particular, whether the neutron star accretes enough mass to 
force it to collapse to a black hole prior to loss of the 
envelope depends entirely on the subsequent evolution of a
thick, advection dominated, accretion disk. Little is known 
about the behavior of such disks, opening plenty of room 
for legitimate dispute as to the outcome.

Simple estimates suggest that the disks formed in the simulations 
are sufficiently hot and thick to support accretion rates which 
could be as large 
as the inferred Bondi-Hoyle rate in the common envelope phase, 
provided only that the disk viscosity is not surprisingly small. 
The presence of a disk does not create a serious bottleneck 
in the accretion flow at $R \sim R_a$. However, more complex and 
uncertain physics is likely to 
come into play close to the neutron star. Jets are observed almost 
universally from accreting systems possessing thick, hot 
accretion disks, of the kind envisaged here. In the case of 
SS433, which arguably is the closest prototype to the physical 
condition of a neutron star in a common envelope, the mass 
outflow in the jets is extremely strong (e.g. Watson et al. 1986).
During common envelope evolution, a jet would allow for a strong feedback 
of energy from close to the neutron star surface into the 
stellar envelope, leading to a more rapid ejection of the 
envelope than would be possible from gas and gravitational 
drag alone. Since the fate of the neutron star depends on 
a sensitive balance between the rate at which it accretes 
and the rate at which energy is deposited into the common 
envelope (Brown 1995), this would improve the 
chances of neutron star survival both by lowering the 
accretion rate and by reducing the epoch of common 
envelope evolution. In general, jets must be avoided 
at all costs if the neutron star is to be able to accrete 
a large mass during inspiral. It is also possible that enough of the 
radiated neutrino energy could be absorbed at a larger radius to drive 
explosions, as discussed in the context of spherical accretion models by
Fryer, Benz \& Herant (1996).

Observationally, several binary pulsars are known whose 
properties would be consistent with the neutron star having 
survived a phase of common envelope evolution. Camilo et al. (1996) 
identify four pulsars which have relatively large companion masses 
(in excess of 0.45 $M_\odot$), and which do not follow the 
eccentricity--orbital period relation expected for lower 
mass binary pulsars (Phinney 1992). These systems are 
likely to have undergone deep common envelope evolution 
(van den Heuvel 1994), although probably the companions 
were of rather lower mass (1--6 $M_\odot$) than we have 
been discussing here. Nonetheless, what is striking is 
that these neutron stars appear to have not only survived, 
but are now observed to be rotating with spin periods that 
are rapid, yet significantly slower than pulsars believed to 
have been spun-up via disk accretion. This is consistent 
with the scenario argued here, in which common envelope 
evolution involves accretion rates that are vastly 
super-Eddington yet still insufficient to force collapse 
of the neutron star to a black hole. It is also 
consistent with the suggestion of Brown, Lee \& Bethe (1999), who 
point out that the inferred masses of black holes in transient 
sources (which appear to be a large fraction of the mass of the 
immediate progenitor) imply that hypercritical accretion onto 
black holes in supernovae must be reasonably efficient. In 
supernovae, mass must be physically ejected from the system 
to avoid being accreted eventually, whereas in the common 
envelope case a modestly lowered accretion rate can suffice 
to enable the neutron star to survive until the envelope 
is lost. 

\acknowledgements

PJA thanks Brad Hansen for many useful discussions, and 
Space Telescope Science Institute for their usual hospitality. 
ML acknowledges support from NASA Grant NAG5-6857.

\newpage

\begin{figure}
\plotone{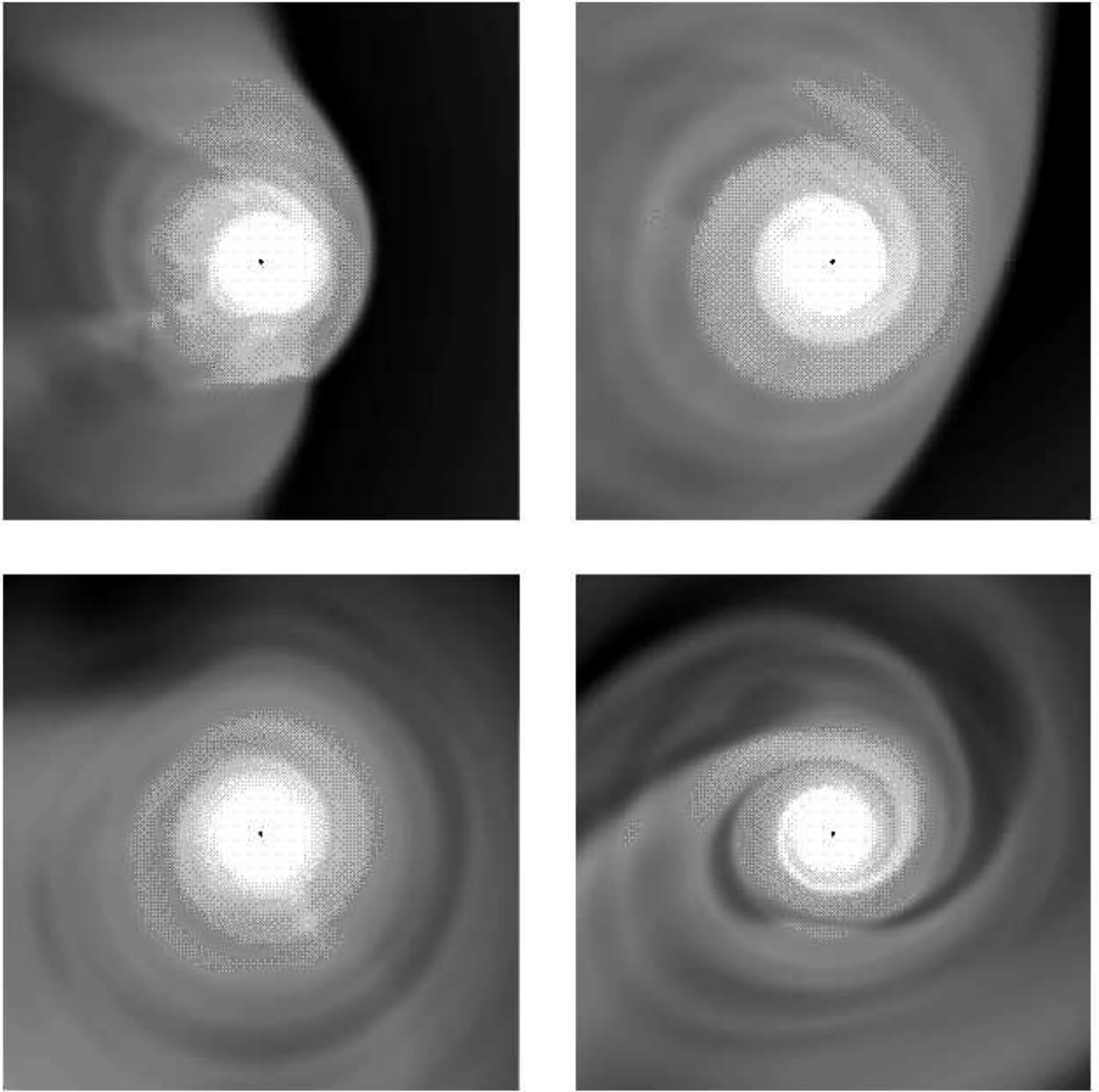}            
\caption{Density (on a logarithmic scale) for different values of 
         the density gradient $\epsilon_\rho$. Top left $\epsilon_\rho = 0$, 
         top right $\epsilon_\rho = 0.1$, lower left $\epsilon_\rho = 0.2$, 
         lower right $\epsilon_\rho = 0.4$. The images measure $4 R_a$ on 
         a side.}
\end{figure}

\begin{figure}
\plotone{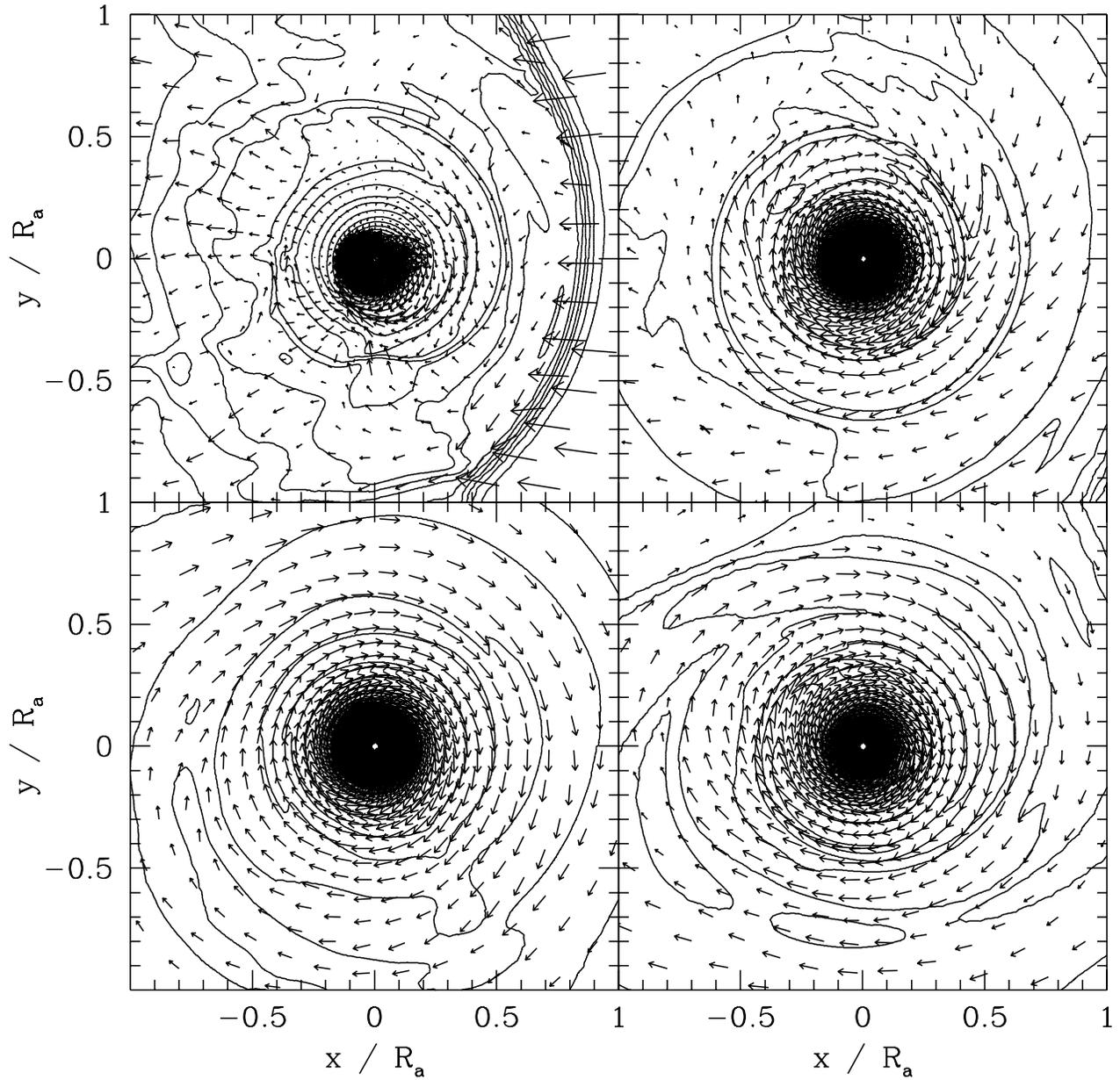}            
\caption{Density and velocity fields in the inner region of the flow. The 
         panels represent the same sequence of increasing $\epsilon_\rho$ 
         as Fig.~1, with contours plotted at $\Delta \log (\rho) = 0.1$. 
         The length of the vectors has been scaled to the local 
         Keplerian velocity $v_k \propto R^{-1/2}$.}
\end{figure}

\begin{figure}
\plotone{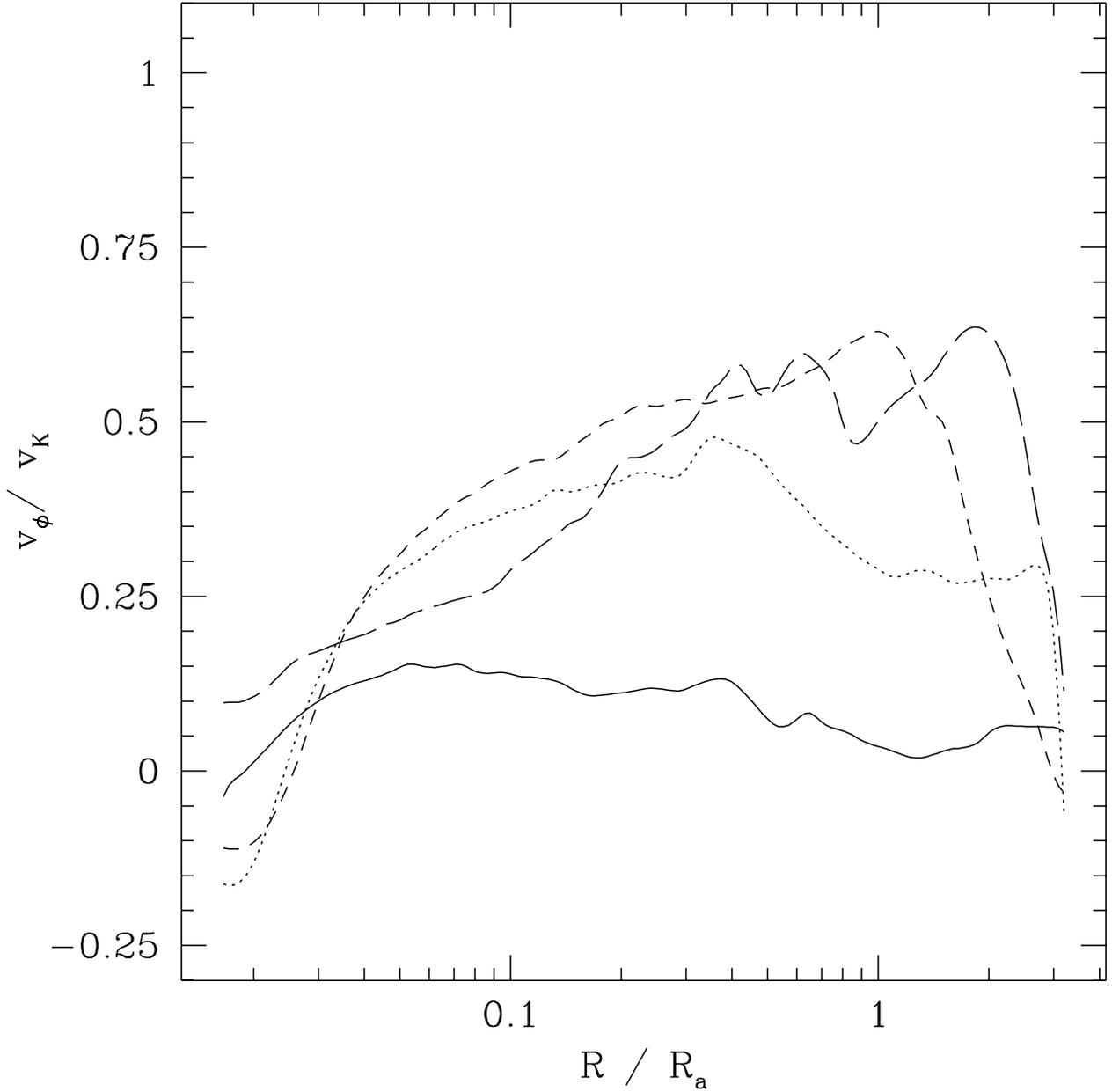}            
\caption{Degree of rotational support for different 
         values of $\epsilon_\rho$. Plotted is the volume averaged 
         rotational velocity $v_\phi$, normalised to the local 
         Keplerian velocity $v_k$, as a function of the radius $R$ 
         in units of the accretion radius $R_a$. The solid line 
         shows results for $\epsilon_\rho = 0$, the dotted line 
         $\epsilon_\rho = 0.1$, the short dashed line $\epsilon_\rho = 0.2$, 
         and the long dashed line $\epsilon_\rho = 0.4$.}
\end{figure}
	
\end{document}